\documentstyle[12pt]{article}
\input epsf
\newcommand{\be}{\begin{equation}}
\newcommand{\ee}{\end{equation}}
\newcommand{\bea}{\begin{eqnarray}}
\newcommand{\eea}{\end{eqnarray}}
\newcommand{\lb}{\label}
\begin{document}
\begin{titlepage}
\title{Direct Detection of Primordial Gravitational Waves in a BSI 
Inflationary Model}
\author{David Polarski\\
\hfill \\
Lab. de Math\'ematique et Physique Th\'eorique, UPRES-A 6083\\
Universit\'e de Tours, Parc de Grandmont, F-37200 Tours (France)\\
\hfill\\
D\'epartement d'Astrophysique Relativiste et de Cosmologie,\\
Observatoire de Paris-Meudon, 92195 Meudon cedex (France)}

\date{\today}
\maketitle

\begin{abstract}
We investigate the possibility for a direct detection by future space 
interferometers of the stochastic gravitational wave (GW) background 
generated during the inflationary stage in a class of viable 
$\Lambda$CDM BSI models. At frequencies around $10^{-3}$Hz,  
maximal values $\Omega_{gw}(\nu)\sim 3\times 10^{-15}$ are found, 
an improvement of about one order of magnitude compared to 
single-field, slow-roll inflationary models. 
This is presumably not sufficient in order to be probed in the near future. 
\end{abstract}

\end{titlepage}

\section{Introduction}
The generation of a primordial gravitational wave background in 
inflationary models is, from a theoretical point of view, an essential
prediction for all of them, it is a genuine quantum gravitational effect. 
Indeed, in addition to a background of scalar perturbations \cite{hsg} of 
quantum origin which, in the inflationary paradigm, 
is responsible for the eventual formation of large scale structure, 
a stochastic background of primordial GW is produced 
during the inflationary stage on a vast range of frequencies from 
$10^{-18}$ up to $10^{10}$ Hz.
In our model, like in many inflationary models, the spectral energy density 
$\Omega_{gw}(\nu)$ is nearly flat in the range 
$10^{-16}{\rm Hz}<\nu<10^{10}$Hz so that
$\sqrt{\rangle h^2 \langle}\propto \nu^{-1}$ for these frequencies. 
Hence, the maximal chances to detect 
this GW background is with space interferometers as they could operate on much 
lower frequencies (larger wavelengths) than ground based detectors, 
typically in the range around $10^{-3}$Hz (see for example, the LISA 
project \cite{Dan95}). 
Remarkably enough, the generation of such a GW background during a de 
Sitter stage and its possible direct detection was first considered 
in a prescient article \cite{Sta79} even before the advent of any concrete 
inflationary scenario and suggested as observational evidence for 
such a stage in the early universe. 
These, and later calculations \cite{Whi92,Lid94,Tur97}, have shown that a 
direct detection, even on the relatively low frequencies $\nu\sim 10^{-3}$Hz  
requires a technological {\it tour de force} to the very least. 

The detection of this gravitational wave background is possible in different 
ways. On very large 
scales they leave an imprint through their contribution to the Cosmic 
Microwave Background (CMB) temperature anisotropy \cite{Ruba82,Sta85}. 
However it is impossible 
to separate the temperature anisotropy multipoles $C^T_{\ell}$ generated by 
the GW from the multipoles $C^S_{\ell}$ generated by the scalar 
perturbations, only the total multipoles $C_{\ell}=C^S_{\ell}+C^T_{\ell}$ 
are measured.   
The GW leave further also a characteristic signature on the CMB polarization 
and this constitutes a promising experimental way, one of the goals set by the satellite mission Planck \cite{Pl}, to discriminate tensorial from scalar 
perturbations. 
The ratio $\frac{C^T_{\ell}}{C^S_{\ell}}$, even 
for small multipole numbers, is usually rather small. 
Aside from the fact that in viable single-field slow-roll 
models the GW contribution to the CMB temperature anisotropy is subdominant, 
the heighth of the Doppler peak around ${\ell}\sim (200-300)$ 
places now a very stringent constraint in this respect 
\cite{Sal95,Mar96,Teg98,Zi98} (see \cite{LPS98} for a recent discussion).
These can be partially 
evaded in a model with Broken Scale Invariance (BSI) \cite{Sta92,MNRAS98} 
(see, e.g., \cite{ARS97} for different BSI models) 
and a cosmological constant (the best fits correspond to $\Omega_{\Lambda}\sim 
(0.6-0.7)$), in which a GW background can be generated with 
$\frac{C^T_{10}}{C^S_{10}}\leq 1$, still in agreement with observations.
For a sufficiently large amount of GW, polarization measurements can further 
help for a very accurate extraction of the inflationary free parameters 
\cite{LPP98}. These aspects were already considered in earlier work.
As mentioned above, investigation of the possibility for a direct detection 
of the GW stochastic 
background generated during slow-rolling of the inflaton led to rather 
pessimistic conclusions. 
However, as such a detection would open a new observational window, it 
remains an exciting possibility to investigate for inflationary models 
which are {\it not} of the single-field slow rolling type.  
It could give us new information about the inflaton potential, and help 
in particular in further constraining inflationary models. This can have 
a significant impact when one is 
looking for a realistic inflationary model based on high energy physics 
models. 

It is our aim to investigate the prospects for a direct 
detection of the stochastic background of gravitational waves produced in 
a particular inflationary model with Broken Scale Invariance on low 
frequencies $\sim 10^{-3}$Hz presumably to be probed by future space 
interferometers.
This observational aspect of our BSI model was not yet considered 
and will be addressed in this letter. 
We will show that in our BSI models with inverted step, which we recall are 
those models which give the best 
fit to the observations, the prospects for a direct detection of the 
generated primordial GW background are sensibly improved as compared to 
usual single-field slow-roll inflation.
Indeed, the spectral energy density obtained in our BSI model will be shown 
to be about one order of magnitude higher than in single-field slow roll 
inflationary models on frequencies $\sim 10^{-3}$Hz. 
However we stress that even for our model, direct detection of the generated 
GW background remains a very hard experimental challenge.
Nevertheless, our result implies that the GW background generated in some 
inflationary models which are viable regarding observations but not of 
the single-field slow-rolling type, offer a better prospect for direct 
detection on low frequencies $\sim 10^{-3}$Hz, and our BSI model is just 
such an example. We will first review the basic details of our BSI model and 
its primordial GW background.

\section{Primordial quantum gravitational waves in our BSI model}

The primordial GW produced during the inflationary stage originate from 
vacuum fluctuations of the quantized tensorial metric perturbations. Each 
polarization state $\lambda$ -- where $\lambda=\times,+$, and the
polarization tensor is normalized to $e_{ij}({\bf k})~e^{ij}({\bf k})=1$ -- has an 
amplitude $h_{\lambda}$ (in Fourier space) given by
\begin{equation}
h_{\lambda}=\sqrt{32\pi G}~\phi_{\lambda}
\end{equation}
where $\phi_{\lambda}$ corresponds to a real massless scalar field. The 
production of a GW background is a generic feature of all inflationary models.

We will consider in this letter a particular inflationary model, a BSI 
model in which the inflaton potential $V(\varphi)$ has a rapid change in 
slope from $A_+>0$ to $A_->0$ (when $\varphi$ decreases) in some neighbourhood 
$\Delta\varphi$ of $\varphi_0$ \cite{Sta92}.
Near the point $\varphi_0$, the second slow-roll condition
$|V''|\ll 24\pi GV$ is violated, while the first slow-roll condition
$V'^2\ll 48\pi GV^2$ is still valid. 
This is why the scalar perturbation spectrum $k^3\Phi^2(k)$ is non-flat around
the scale $k_0=a(t_{k_0}) H_{k_0}$, which crosses (out of) the 
Hubble radius when $\varphi(t_{k_0})=\varphi_0$, ($H\equiv \dot{a}/a$ is 
the Hubble parameter). The spectrum has basically a ``step'' structure and we 
have in the neighbourhood of the step 
\bea
k^3 \Phi^2(k) &=& \frac{81}{50}\frac{H_{k_0}^6}{A_+^2}~~~~~\varphi >\varphi_0\\
&=& \frac{81}{50}\frac{H_{k_0}^6}{A_-^2}~~~~~\varphi <\varphi_0~, \label{Phi}
\eea
where $H_{k_0}$ is the value of the Hubble parameter at 
$\varphi =\varphi_0$, 
$H_{k_0}^2=\frac{8\pi G}{3}V(\varphi_0)$.
We see in particular from eq.(\ref{Phi}) that 
an inverted step is obtained for $p<1$.
The shape of the spectrum is solely determined by the parameter 
$p\equiv \frac{A_-}{A_+}$ and independent of the characteristic scale $k_0$.
This form is further universal as it does not depend on $\Delta\varphi$ 
nor on the details of $V(\varphi)$ inside $\Delta\varphi$.     
In order to study concrete inflationary models, we have to make additional 
assumptions about the inflaton potential $V(\varphi)$.
We will assume here that the inflaton potential 
satisfies the slow-roll conditions far from the point $k_0$ and we will 
also assume a particular behaviour of the spectral indices $n_T(k)$ and 
$n_s(k)$. For more details 
and for analytic expressions of the power spectrum, we refer the interested 
reader to \cite{Sta92,MNRAS98}. 

In this model, for fixed normalization and shape of the spectrum, a GW 
background of arbitrary amplitude can be produced. It was shown that a large 
contribution of this GW background to the CMB anisotropy 
(for $\ell<80$) is possible and compatible 
with the observations even though slow-roll regime is assumed on very large 
scales ($k<k_0$) \cite{LPS98}. This is because of the possibility to have an 
inverted step in the scalar fluctuations power spectrum $k^3\Phi^2(k)$, which 
is in turn rendered possible by the presence of a cosmological constant.   
Note that the rise of the multipoles $C_{\ell}$ on the range 
$50\leq \ell\leq 100$, precludes a higher contribution of the GW. 
Finally, the characteristic scale $k_0=0.016$ is fixed in order to agree with 
a potential feature in the matter power spectrum on scales 
$\sim 125~h^{-1}$Mpc. 
We will now investigate whether this GW background, which has acquired 
nowadays the properties of a classical stochastic background \cite{CQG96}, 
can be detected in the frequency range around 0.001 Hz.

\section{Direct detection of the primordial GW}
Regarding the direct detection of the GW background, special interest is 
attached to the quantity 
$\Omega_{gw}(\nu)\equiv \frac{1}{\rho_{cr}}\frac{d \rho_{gw}}{d \ln \nu}$
where $\rho_{cr}=\frac{3H_0^2}{8\pi G}$ is the critical energy density 
($H_0$ is the Hubble parameter today not to be confused with $H_{k_o}$), 
$\rho_{gw}$ is the energy density of the GW background
while $\nu$ is the frequency of the GW. A direct detection of primordial GW 
on very large scales comparable to the Hubble radius is clearly hopeless 
because they correspond to exceedingly low frequencies $\sim 10^{-18}$ Hz. 
The prospects for a direct detection are evidently better for space detectors 
which could operate on frequencies $\nu \sim 10^{-3}$ Hz. For single-field 
slow-roll 
inflation the quantity 
$\Omega_{gw}(\nu=10^{-3}{\rm Hz})\sim 10^{-15}-10^{-16}$ 
\cite{Tur97}, which is very low in order to be detected.
Drastic improvement in the models predictions is needed, putting aside 
possible upgrading in sensitivity of future space GW detectors.  
When slow-roll is assumed on large scales, a higher value of 
$\frac{C^T_{10}}{C^S_{10}}$ will have two competing consequences. On one hand, 
it implies a higher value of the Hubble parameter on very large scales 
and therefore a higher overall normalization of the GW's power spectrum. On 
the other hand however, the latter will also decrease quickier towards small 
scales (large $k$'s). 
Hence a related question is: for given value of 
$\Omega_{gw}(\nu)$, what is the corresponding value of the quantity 
$\frac{C^T_{10}}{C^S_{10}}$~? For single-field slow-roll inflation we have 
\cite{PLB95}   
\be
\frac{C^T_{10}}{C^S_{10}} \equiv \frac{\langle |a_{10m}|^2\rangle_{gw}}
{\langle |a_{10m}|^2\rangle_{ap}} = {\cal K}_{10} |n_T| \sim 5~|n_T|. \lb{TS} 
\ee
where the quantity ${\cal K}_{10}$ depends on the cosmological parameters. 
In other words, how is direct detection of the GW related to their indirect 
detection through the CMB anisotropy which they produce on large scales 
($\ell\sim 10$)?
We will show that in our BSI model the prospects for a direct detection are 
sensibly improved and that 
a ratio $\frac{C^T_{10}}{C^S_{10}}$ is obtained which is, for given value 
of $\Omega_{gw}(\nu)$, significantly higher than in models where slow-roll 
is assumed {\it everywhere}.  

A detailed numerical study has shown that BSI models with an inverted step 
with a ratio $\frac{C^T_{10}}{C^S_{10}}$ as high as unity \cite{LPS98} are 
viable. This is the crucial property of our model which does improve the 
possibility for a direct detection in two ways. 
First, the presence of a cosmological constant combined with an inverted step 
for the adiabatic perturbations power spectrum allows, as mentioned above, a 
fairly high ratio $\frac{C^T_{10}}{C^S_{10}}$, and hence, on very large 
scales probed by the CMB anisotropy, a value for the 
Hubble parameter which is higher than in usual single field slow-roll 
inflationary models. We remind that in our model too, the slow-roll equations 
are valid on those scales corresponding to $k<k_0$.
In the second 
place, an inverted step corresponds to values $p<1$, hence at the break 
point there will be a corresponding jump of the tensorial spectral index 
$n_T$, it will be larger behind the scale $k_0$.
Indeed, in the model investigated in ~\cite{MNRAS98,LPS98,LPP98}, there is 
a jump in the first derivative of the inflaton potential and a corresponding 
jump in the (smoothed) tensor spectral index $n_T$ with:
\be
|n_T(k_0^-)|=p^2~|n_T(k_0^+)|,\label{n}
\ee 
$k_0^+$, resp. $k_0^-$, refers to $k$ values smaller, resp. larger, than 
$k_0$ in the vicinity of $k_0$. 
  
If we assume that the universe went after the 
inflationary phase through a radiation dominated stage (eventually followed 
by a matter-dominated stage), the following result is obtained for a GW that 
reentered the Hubble radius deep inside the 
radiation dominated stage, though still first crossed the Hubble radius 
long before the end of the inflationary phase (with $\hbar=c=1$) 
\be
\Omega_{gw}(\nu)= \frac{2}{3\pi}~\frac{g}{2}~\Omega_{\gamma}
~{\tilde H}_k^2\label{Ome} 
\ee
The densities (relative to the critical density) $\Omega_X$ are evaluated 
today. In eq.(\ref{Ome}), $\Omega_{\gamma}$ refers for the CMB photons and 
$g$ is the total effective number of degrees of freedom of 
all relativistic matter at $t=t_{eq}$, the time when relativistic and 
non relativistic matter have equal energy densities, 
finally ${\tilde H}_k$ is the Hubble parameter, expressed in Planck mass 
units, at Hubble radius crossing during inflation for that scale 
corresponding today to a GW with frequency $\nu$. 
In eq.(\ref{Ome}), $g$ plays the role of an additional cosmological parameter. 
In the assumption that all relativistic matter stays relativistic from 
$t_{eq}$ till today, eq.(\ref{Ome}) corresponds to the sometimes quoted result 
$\frac{2}{3\pi}~{\tilde H}_k^2~(1+z_{eq})^{-1}$ multiplied by $\Omega_{NR}$, 
where $\Omega_{NR}$ refers to all non-relativistic matter. 
For $\Omega_{NR}\sim 0.3$, which corresponds to our $\Lambda$CDM BSI models, 
eq.(\ref{Ome}) gives therefore a sensibly {\it smaller} $\Omega_{gw}(\nu)$.    
Finally, for frequencies of interest here, eq.(\ref{Ome}) is completely 
insensitive to a value of $\Omega_{\Lambda}$ as large as 
$\Omega_{\Lambda}\sim 0.7$ (as recent observations suggest).

If we assume that the inflationary phase took place at very high energies 
in a pure radiation dominated universe, the result simplifies to
\be
\Omega_{gw}(\nu)= \frac{2}{3\pi}~{\tilde H}_k^2~.\label{Ome1}
\ee 
In order to compare with accurate numbers, eq.(\ref{Ome}) is conveniently 
recast into 
\be
\Omega_{gw}(\nu)~h^{2}_{60} = 2.44~\frac{g}{3.36}~H^{*2}_k
~\times 10^{-15},\label{Ome2} 
\ee 
The value $g=3.36$ corresponds to the case of relativistic matter 
consisting of photons plus three flavours of relativistic neutrinos and 
$H^{*2}_k\equiv {\tilde H}_k^2\times 10^{10}$.
Though we will take $g=3.36$ in our calculations, a different value is an 
interesting possibility.
The quantity ${\tilde H}_k$ depends on the inflaton potential on all 
scales $k>k_0$, this 
behaviour is not specified by the presence of a break at $k=k_0$. 
${\tilde H}_k$ is further model-dependent and depends therefore on all 
parameters, both cosmological and inflationary.
We have 
chosen $h_{60}\equiv \frac{H_0}{60}$, rather than say $h_{50}$, because the 
best viable BSI models have $h$ closer to $h=0.6$ and also because of current  
estimates. Finally, for those models in good agreement with observations 
which have a large contribution of the GW background to the CMB anisotropy, 
we have $H^{*2}_k\sim (1-2)$.
Two simple examples for the running of $n_T$ are provided by: a) $ n_T = 
n_s-1$ = const; b) $ n_s = 1$.   
These two behaviours were already considered in 
\cite{LPS98} on large scales, $k<k_0$. We will now also assume that they are 
valid on scales $k>k_0$.  
The second case is also a very good approximation when $n_s$ is very close 
to one. As the best values are evidently obtained for the second case, we 
will give here estimates for this case only. 
In the most favourable cases, corresponding to the cosmological parameters 
$\Omega_{\Lambda}\sim 0.65,~h\sim 0.6$, and to the inflationary parameters 
$p\sim 0.6,~n_T(k^+_0)\sim -0.12$ (see Table 1), our BSI model yields 
\be
\Omega_{gw}(\nu=10^{-3}{\rm Hz})\sim 2.5 \times 10^{-15}\label{V}
\ee 
This is to be compared, e.g., with the result
$\Omega_{gw}h^2(\nu=10^{-3}{\rm Hz})\approx 3.5\times 10^{-16}$ 
obtained in \cite{Tur97} for $h=0.6,~g=3.36$ ($\Omega_{\Lambda}=0$), 
$n_T\approx -0.025$.  

As noted earlier, a large GW contribution to the CMB fluctuations, as it 
implies a large spectral index, will yield a low value for 
$\Omega_{gw}(\nu=10^{-3}{\rm Hz})$. 
This is less constraining in our model due to the jump in $n_T$, eq.(\ref{n}). 
Hence, models yielding a value as high as (\ref{V}), admit simultaneously a 
relatively high contribution of the GW to the CMB anisotropy with
$\frac{C^T_{10}}{C^S_{10}}\sim (0.8-0.9)$.
Note that agreement with the observations studied earlier in \cite{LPS98} 
implies a correlation between the cosmological parameters $h$ and 
$\Omega_{\Lambda}$. For example, for $h=0.6,~\Omega_{\Lambda}=0.7$, slightly 
higher values $\Omega_{gw}\approx 3\times 10^{-15}$ are obtained, however 
these models are in poor agreement with the CMB anisotropies.  
Higher values of $\Omega_{\Lambda}$ will therefore 
correspond to higher values of $h$. Then however, the main effect will be a 
decrease of $\Omega_{gw}$ when higher values of $h$ are considered.

The value we get for $\Omega_{gw}(\nu=10^{-3}{\rm Hz})$ represents a 
substantial improvement of nearly one order of magnitude as compared to 
usual single-field, slow-roll inflation. Though a sensitivity on such a 
level remains a technological challenge, these results show that the 
prospects for a direct detection of the GW background generated in realistic 
inflationary models can be improved compared to usual single-field 
slow-roll inflation. We stress again 
that the values (\ref{V}) are obtained for BSI inflationary models which are 
observationally viable and which satisfy the slow-roll equations on scales 
where the perturbations power spectrum gets normalized. 
Also, these models corespond to a minimal change of the usual standard 
inflationary scenario, in contrast to more unconventional inflationary 
scenarios like e.g. the PBB (Pre Big Bang) scenario \cite{Ve}, where a direct 
detection is possible due to a significant rise in $\Omega_{gw}$ on much 
higher frequencies accessible to ground based detectors \cite{Li99},
but not on the lower frequencies around $10^{-3}$Hz considered here.

It was already realized in the literature that the constraints for a direct 
detection of primordial GW might be loosened when the slow-roll assumption 
is relaxed. This is indeed the case in our BSI models. However not all our BSI 
models will improve the prospects for direct detection but only those with 
inverted step which were previously shown to give the best fit to the 
observations. In 
this sense, we stress that the value of the inflationary parameter $p$ is
not introduced ad hoc in order to improve direct GW detection but that it 
comes from completely independent observational constraints.  
In conclusion, our results show that the sensitivity level required by our 
model is nearly one order of magnitude lower than required by usual 
single-field slow roll inflationary models. 
Whether this will be sufficient for detection 
by some future experiment is a fascinating technological problem on its own, 
however one beyond the scope of this letter.
As the single-field slow roll  models are not necessarily ruled out in 
the presence of a large  cosmological constant, this fact constitutes a clear 
distinction between them. A positive detection, if technological progress 
allows it in the future, could be one more hint at inflationary models, like 
the specific $\Lambda$CDM BSI models considered here, which are not of the 
simplest, single-field, slow-rolling type. 

\vspace{1cm}
\noindent

{\bf Acknowledgements}
\par\noindent
It is a pleasure to thank A.A. Starobinsky for stimulating discussions.

\vfill\eject
\begin{table}
\begin{center}
\begin{tabular}{|r|l|l|l|l|l|}
\hline
 ${\rm cosm.}\backslash {\rm infl.}$ & $~p~$ & $|n_T|$ & $H^*_{k_0}$ & $~R~$  & $\Omega_{gw}$     \\
\hline
$\Omega_{\Lambda}=0.60$   & .92 &.001  & 0.03  &.004  & 0.08  \\
\cline{2-6}
$h=0.60$                  & .72 &.1    & 2.16  &.61  & 1.97  \\
\hline
$\Omega_{\Lambda}=0.65 $  & .79 &.001  & 0.03  &.004  & 0.08  \\
\cline{2-6}
$h=0.60$                  & .74 &.025  & 0.76  &.12  & 1.30  \\
\cline{2-6}
                          & .58 &.125  & 2.31  &.85  & 2.41  \\
\cline{2-6}
                          & .57 &.13   & 2.33  &.91  & 2.40  \\
\hline
$\Omega_{\Lambda}=0.70$   & .65 &.1    & 2.1   &.60  & 1.61  \\
\cline{2-6}
$h=0.70$                  & .58 &.135  & 2.31  &.97  & 1.68  \\
\hline
$\Omega_{\Lambda}=0.75$   & .52 &.1    & 2.06  &.60  & 1.99  \\
\cline{2-6}
$h=0.70$                  & .47 &.13   & 2.25  &.91  & 2.09  \\
\hline
\end{tabular}
\end{center}
\caption{Table 1 displays models for different cosmological 
parameters $\Omega_{\Lambda},~h$, and inflationary parameters (see text) 
$p,n_T\equiv ~n_T(k_0^+),~H^*_{k_0},~R\equiv \frac{C^T_{10}}{C^S_{10}},
~\Omega_{gw}\equiv \Omega_{gw}(\nu=10^{-3}{\rm Hz})\times 10^{15}$. 
All these models are in agreement with observations. $n_S=1$ is assumed on 
scales $k>k_0,~k<k_0$. $\Omega_{gw}$ can be about one order of magnitude 
higher than in single-field slow-roll inflationary models.}
\end{table}

\end{document}